\documentclass[useAMS,usenatbib]{mnras}
\usepackage{graphicx}
\usepackage{dcolumn}
\usepackage{bm}
\usepackage[usenames]{color}

\usepackage{amsmath}
\usepackage{amssymb}

\DeclareMathOperator{\arcosh}{arcosh}

\title[The Hubble stop radius and the galaxy group mass]{The radius, at which a galaxy group stops the Hubble stream, and the group mass: an exact analytical solution}

\author[A. N. Baushev]{A. N. Baushev$^{1}$\\
	$^{1}$Bogoliubov Laboratory of Theoretical Physics, Joint Institute for Nuclear Research\\
	141980 Dubna, Moscow Region, Russia}
\begin{document}

\date{}

\pagerange{\pageref{firstpage}--\pageref{lastpage}} \pubyear{2019}

\maketitle

\label{firstpage}

\begin{abstract}
The gravitational field of a galaxy group or cluster slows down the Hubble stream and turns it speed to zero at some radius $R_0$. We offer an exact analytical relation between $R_0$ and the mass of the group.
\end{abstract}


\section{Introduction}
Considering a group (or a cluster)\footnote{All our reasoning is equally applicable to groups and clusters of galaxies, and hereafter we will name them 'groups' for the sake of shortness.} of galaxies, one may see that the radial velocities of distant galaxies with respect to the group center grow linearly, as it should be in the Friedmann's Universe. However, as we approach the group, average radial velocity drops faster than linearly and reaches zero at some radius $R_0$ (as a good illustration for the for the Local Group instance, see Figure~1 in \citep{makarov2009}). The reason is obvious: the group is an overdensity, and the additional gravitational field of it tends to stop the Hubble flow\footnote{$R_0$ is sometimes named "the turnaround radius". This name is discussable: generally speaking, nothing turns around at this radius. We call $R_0$ "the stop radius", though it is only a question of terminology.}. The value of $R_0$ can be measured in astronomical observations, and this is one of methods to determine the group mass $M$ (to be more precise, the mass inside $R_0$). However, one needs the relationship between $M$ and $R_0$.

Curiously, there are several approaches to the problem in literature. Old papers (for instance, \citep{olson1979, lb1981, giraud1986}) considered the problem analytically, but the articles were written in the time when the cosmological constant was believed to be negligible. Some authors (e.g., \citep{makarov2009}) use approximate formulas for $M(R_0)$. Now N-body simulations are frequently performed to obtain $M(R_0)$ (for example, \citep{hanski2001, penarrubia2014}). Of course, N-body simulations is a power method allowing to consider, e.g.  realistic non-spherical models of the Local Group \citep{penarrubia2014}. On the other hand, the N-body simulations have their own weak points \citep{17, 21}, while the spherically-symmetric model is, in many instances, quite sufficient to describe observational data. Strictly speaking, the stop radius $R_0$ may be introduced only for a spherically-symmetric system: the stop surface of a strongly asymmetric group is essentially not a sphere. 

Meanwhile, the exact relationship $M(R_0)$ for a spherically-symmetric system is quite simple and can be found analytically; therefore, expediency of usage of some approximate equations or simulations in this case seems questionable. The aim of this short letter is to derive the solution of this simple task.

\section{Calculation}
\subsection{A brief cosmological outline}
The metric of a homogeneous isotropic universe can be represented as $ds^2=c^2dt^2-a^2(t) dl^2$, where $dl$ is an element of
three-dimensional length and $a$ is the scale factor of the Universe. We denote the present value of $a$ by $a_0$. One may introduce the \emph{present-day} critical density of the universe $\rho_{c,0}=\dfrac{3H_0^2}{8\pi G}$, where $H_0$ is the present-day value of the Hubble constant. Then the Hubble constant evolution is given by the Friedmann equation (see e.g. \citep[eqn. 4.1]{gorbrub1})
\begin{eqnarray}
&H^2(t)\equiv \left(\dfrac{da}{a dt}\right)^2 = \\
&=\dfrac{8\pi}{3}G \left[\rho_{M,0}\left(\dfrac{a_0}{a}\right)^3\!\!+\rho_{\gamma,0}\left(\dfrac{a_0}{a}\right)^4\!\!+\rho_{\Lambda,0}+\rho_{a,0}\left(\dfrac{a_0}{a}\right)^2\right].\nonumber \label{24a1}
\end{eqnarray}
Here $\rho_{M,0}$, $\rho_{\gamma,0}$, $\rho_{\Lambda,0}$, $\rho_{a,0}$ are the \emph{present-day} matter, radiation, dark energy, and curvature densities of the Universe, respectively. The curvature density is defined by $\dfrac{8\pi}{3 c^2} G \rho_{a,0} = \dfrac{k}{a_0^2}$, where $k=-1, 1, 0$ if the universe density is higher, smaller, or equal to the critical one, respectively. It is common practice to use the ratios of the present-day densities of the universe components to $\rho_{c,0}$: $\Omega_{M,0}\equiv \rho_{M,0}/\rho_{c,0}$ etc\footnote{The values $\Omega_{M,0}$, $\Omega_{\Lambda,0}$ etc. are frequently denoted simply by $\Omega_{M}$, $\Omega_{\Lambda}$ etc. We use the additional sub-index to emphasize that we imply the present-day values.}. Then (\ref{24a1}) may be rewritten as 
\begin{equation}
\frac{da}{a dt} = H_0 \sqrt{\Omega_{M,0}\left(\frac{a_0}{a}\right)^3\!\!+\Omega_{\gamma,0} \left(\frac{a_0}{a}\right)^4\!\!+\Omega_{\Lambda,0}+\Omega_{a,0}\left(\frac{a_0}{a}\right)^2}. \label{24a2}
\end{equation}
Here we have assumed that the dark energy may be described by a non-zero cosmological constant (i.e., $p_\Lambda=-\rho_{\Lambda,0}=\it{const}$).

We perform our calculations under the following assumptions:
\begin{enumerate}
	\item \label{as1}  The Universe is flat ($\Omega_{a,0}=0$) in absence of structures, and the dark energy may be described by the cosmological constant. Both the assumptions are suggested by modern cosmological observations as the most probable \citep{pdg18}. Moreover, we neglect the radiation term. Now $\Omega_{\gamma,0}\sim 10^{-4}$, and, though the radiation contribution was much larger in the early Universe, the relative error of the group mass determination caused by the disregarding of radiation is also $\sim 10^{-4}$, as we will show.
	\item \label{as2} The size of the group is large with respect to its gravitational radius $R_0\gg R_g\equiv{2GM}/{c^2}$ and small with respect to $c/H_0$.  For real groups $R_0/R_g>10^4$, $R_0 H_0/c<10^{-3}$, i.e., both the conditions are well satisfied. The significance of this assumption will be explained below.
	\item \label{as3} The group of galaxies is spherically symmetric and does not experience any tidal perturbations. Typically, this assumption is not quite valid for real galaxy groups. However, it allows to find a precise analytical solution even in the nonlinear regime, while the accuracy of the astronomical determination of $R_0$ is frequently not that high, and we may neglect the influence of tidal effects and nonsphericity on the mass estimation. Finally, the introduction of the stop radius $R_0$ implies some spherical symmetry of the system: the stop surface of a strongly asymmetric group is not a sphere.
	\item We measure $R_0$ at present (or, to put it differently, the galaxy group has small redshift $z$): strictly speaking, the relationship $M(R_0)$ depends on $z$. We will consider the case when $z\ne 0$ at the end of this letter.
	\end{enumerate}
For a start, we consider our Universe without perturbations. It follows from assumption \ref{as1} that $\Omega_{a,0}=0$, $\Omega_{\gamma,0}\simeq 0$. Then 
\begin{equation}
\Omega_{\Lambda,0}+\Omega_{M,0}=1, \label{24a3}
\end{equation}
and we may integrate (\ref{24a2}), obtaining the well-known equation (see e.g. \citep[eqn. 4.29]{gorbrub1}) for the Universe age $t_0$
\begin{eqnarray}
t_0=H_0^{-1} \frac{2}{3\sqrt{\Omega_{\Lambda,0}}}\ln\left[\dfrac{\sqrt{1-\Omega_{M,0}}+1}{\sqrt{\Omega_{M,0}}}\right]=\\ \nonumber 
=H_0^{-1} \frac{2}{3\sqrt{\Omega_{\Lambda,0}}}\arcosh\left(1/\sqrt{\Omega_{M,0}}\right) \label{24a4}
\end{eqnarray}

\subsection{The function $M(R_0)$}
Now consider the primordial perturbation that later transformed into the galaxy group under review, following the standard method offered in \citet{tolman1934, bondi1947}. We have assumed that there are no tidal effects, i.e., the perturbation may be treated as the only in the whole Universe. Let us consider a sphere of radius $r$ ($r>R_{vir}$) around the group center and discuss its temporal evolution $r(t)$. The principle fact is that the density distribution inside $r$, as well as the matter outside $r$, do not affect the temporal evolution of the sphere $r(t)$ at all, if the system is spherically symmetric. A detailed proof of this statement may be found, for instance, in \citet[chapters 1, 4]{zn2}, but it is pertinent to give a brief outline of it here. 

\begin{itemize}
\item The only component with pressure we have in our system is the dark energy. However, its density and pressure remain constant
in time and space. There is no pressure gradient or jump, in particular, at $r$.
\item A spherically symmetric matter distribution outside some radius $r$  does not create gravitational field inside this radius in the general theory of relativity (GTR), as well as in the Newtonian theory  \citep[chapters 1]{zn2}.
\item Gravitational force created by a spherically symmetric system at some radius does not depend on the matter distribution inside this radius (if the distribution does not violate the spherical symmetry) in the Newtonian gravity. Generally speaking, it is not true in GTR, but in our case (see assumption~\ref{as2}) the deviations may occur only because the pressure also creates gravitational field in GTR, and the pressure depends on the matter distribution. However, the only component with pressure  in our instance (the dark energy) is distributed uniformly, and if we redistribute the pressure-less matter inside $r$ uniformly, it will not affect the gravitational force acting on the particles at $r$ \citep{tolman1934, bondi1947}.
\end{itemize}

Thus, we may assume that the matter inside $R$ is distributed uniformly; even if it is not really so, it does not affect the dependence $R(t)$. Then the spherically symmetric and uniform system inside $r$ may again be described by the Friedmann equation (\ref{24a1}), but with different values of $\rho_{c,0}$, $\rho_{M,0}$ etc. We choose $r=R_0$ and consider its evolution with time $R(t)$, i.e., we consider the previous evolution of the sphere that now stops at the distance $R_0$ from the group center. From (\ref{24a1}) we have
\begin{eqnarray}
 \left(\dfrac{dR}{R dt}\right)^2 =\dfrac{8\pi}{3}G \left[\sigma_{M,0}\left(\dfrac{R_0}{R}\right)^3\!\!+\sigma_{\Lambda,0}+\sigma_{a,0}\left(\dfrac{R_0}{R}\right)^2\right]. \label{24a5}
\end{eqnarray}
Here we have neglected the radiation component, as we did deriving (\ref{24a4}); $\sigma_{M,0}$ is the average matter density inside $R_0$. Of course, $\sigma_{M,0}>\rho_{M,0}$, the galaxy group forms from an overdensity. $\sigma_{\Lambda,0}$ is the dark energy density, and $\sigma_{\Lambda,0}=\rho_{\Lambda,0}$. Contrary to the case of the flat Universe, the curvature density in the perturbation $\sigma_{a,0}\ne 0$. Since the sphere under consideration stops at $R=R_0$, the right part of (\ref{24a5}) turns to zero if $R=R_0$. So
\begin{equation}
\sigma_{M,0}+\sigma_{\Lambda,0}+\sigma_{a,0}=\sigma_{M,0}+\rho_{\Lambda,0}+\sigma_{a,0}=0. \label{24a6}
\end{equation}
We may introduce 
\begin{equation}
\alpha=\dfrac{\sigma_{M,0}}{\sigma_{\Lambda,0}}=\dfrac{\sigma_{M,0}}{\rho_{\Lambda,0}}=\dfrac{\sigma_{M,0}}{\rho_{c,0}\: \Omega_{\Lambda,0}} \label{24a7}
\end{equation}
and substitute the value (\ref{24a6}) for $\sigma_{a,0}$ to (\ref{24a5}). We obtain
\begin{eqnarray}
\left(\dfrac{dR}{R dt}\right)^2 =\dfrac{8\pi}{3}G\rho_{\Lambda,0} \left[\alpha\left(\dfrac{R_0}{R}\right)^3\!\!+1-(\alpha+1)\left(\dfrac{R_0}{R}\right)^2\right]. \label{24a8}
\end{eqnarray}
In order to obtain the time $t_0$ of the sphere expansion from $R=0$ to $R=R_0$, we should integrate (\ref{24a8}). It is convenient to use $x\equiv R/R_0, x\in[0,1]$ as the integration variable. Moreover, we should take into account that 
\begin{equation}
\dfrac{8\pi}{3}G\rho_{\Lambda,0}=\dfrac{8\pi}{3}G\rho_{c,0} \Omega_{\Lambda,0}=H_0^2 \Omega_{\Lambda,0}. \label{24a9}
\end{equation}
Then we obtain
\begin{equation}
t_0=\dfrac{1}{H_0 \sqrt{\Omega_{\Lambda,0}}}\int_0^1\dfrac{dx}{x \sqrt{\alpha/x^3+1-(\alpha+1)/x^2}}. \label{24a10}
\end{equation}

\begin{figure}
	\resizebox{1.1\hsize}{!}{\includegraphics{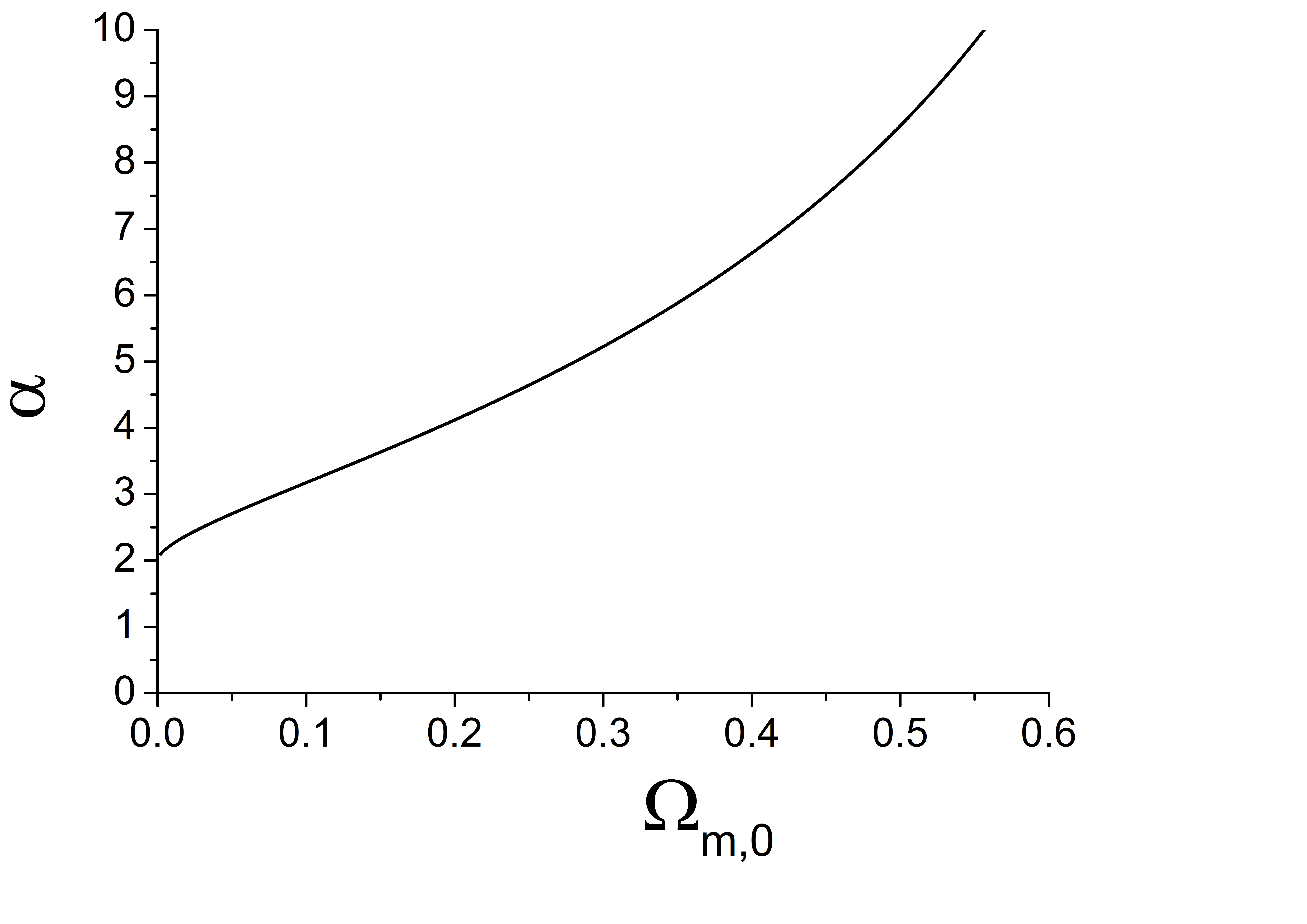}}
	\vspace{-0.7cm}
	\caption{The function $\alpha(\Omega_{M,0})$, implicitly given by equation~(\ref{24a11}) or equations~(\ref{24b1}) and (\ref{24b2}).} \label{24fig1}
\end{figure}

\noindent Of course, the ages of the undisturbed Universe (\ref{24a4}) and the disturbed one (\ref{24a10}) should be the same, since the Universe expansion started simultaneously. Equating (\ref{24a4}) and (\ref{24a10})
\begin{equation}
\frac{2}{3}\ln\left[\dfrac{\sqrt{1-\Omega_{M,0}}+1}{\sqrt{\Omega_{M,0}}}\right]=\int_0^1\dfrac{\sqrt{x}dx}{ \sqrt{\alpha+x^3-x(\alpha+1)}}, \label{24a11}
\end{equation}
we obtain a function $\alpha(\Omega_{M,0})$, binding the average matter density inside $R_0$, $\sigma_{M,0}=\alpha \rho_{c,0} \Omega_{\Lambda,0}$, with the matter fraction in the unperturbed Universe, $\Omega_{M,0}$, which is well known from observations. Equation~(\ref{24a11}) defines $\alpha(\Omega_{M,0})$ implicitly (see figure~\ref{24fig1}) and can be slightly simplified. The integral in the right side of (\ref{24a11}) 
\begin{equation}
I(\alpha)\equiv \int_0^1\dfrac{\sqrt{x}dx}{ \sqrt{\alpha+x^3-x(\alpha+1)}}, \label{24b1}
\end{equation}
is elliptical and cannot be expressed in elementary functions. However, we may simplify equation~(\ref{24a11}) to
\begin{equation}
\Omega_{M,0}=\dfrac{1}{\cosh^2\left(3 I(\alpha)/2\right)}. \label{24b2}
\end{equation}
The integral (\ref{24b1}) is meaningful only if $\alpha>2$. It is not a surprise: if $\alpha=2$, $\sigma_{\Lambda,0}=2 \sigma_{M,0}$. However, the effective gravitational repulsion created by the cosmological constant is two times stronger than the gravitational attraction created by normal matter of equal density \citep{tolman1934}. Thus, if $\alpha\le 2$, the overdensity of matter is just too low to surpass the repulsion of dark energy and stop the Hubble flow, and there is no $R_0$ in this case.

The method of derivation of equation (\ref{24a11}) justifies our neglect of the radiation term $\rho_{\gamma,0}$. Now $\Omega_{\gamma,0}< 10^{-4}$ \citep{gorbrub1}. Radiation dominated in the early Universe, but that epoch was quite short ($<5\cdot 10^5$ years, i.e., less than $10^{-4}$ of the universe age which we equated in (\ref{24a11})). We conclude that the relative error occurring from the neglect of the radiation term does not exceed $10^{-4}$. The accuracy of $R_0$ determination for real systems is, unfortunately, much lower.

\begin{figure}
	\resizebox{1.1\hsize}{!}{\includegraphics{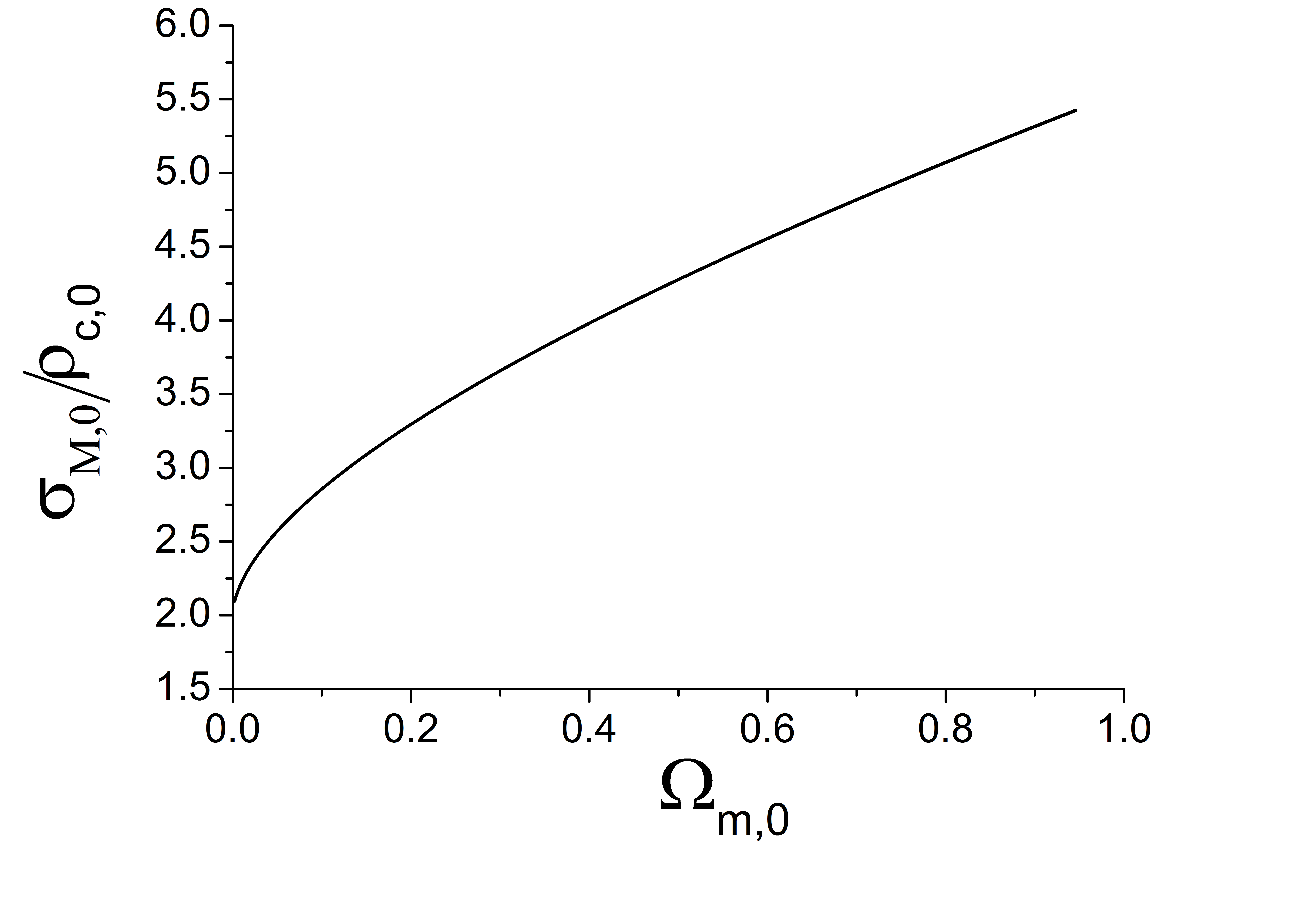}}
	\vspace{-0.7cm}
	\caption{The ratio of the average matter density $\sigma_{M,0}$ inside $R_0$ to the critical density $\rho_{c,0}$ of the Universe, as a function of $\Omega_{M,0}.$} \label{24fig2}
\end{figure}

Function $\alpha(\Omega_{M,0})$ directly gives us the average matter density $\sigma_{M,0}$ inside $R_0$: $\sigma_{M,0}=\alpha(\Omega_{M,0})\sigma_{\Lambda,0} =\alpha(\Omega_{M,0})(1-\Omega_{M,0})\rho_{c,0}$. While $\alpha\to \infty$ if $\Omega_{M,0}\to 1$, the ratio $\sigma_{M,0}/\rho_{c,0} =\alpha(\Omega_{M,0})(1-\Omega_{M,0})$, of course, remains finite. One may obtain directly from equation~(\ref{24a11}) by a limiting process: 
\begin{equation}
\frac{\sigma_{M,0}}{\rho_{c,0}}=\left(\frac{3\pi}{4}\right)^2, \quad \text{if} \quad \Omega_{M,0}=1 \label{24c1}
\end{equation}
Figure~\ref{24fig2} shows the ratio of the average matter density inside $R_0$ to the present-day critical density of the Universe. Assumption~\ref{as2} means that the three-dimensional space inside $R_0$ is almost euclidean (that is why we set assumption~\ref{as2}), and the total matter mass inside $R_0$ is (we substitute equation for $\rho_{c,0}$) 
\begin{eqnarray}
\label{24a12}
&M=\frac43\pi R_0^3 \alpha(\Omega_{M,0})(1-\Omega_{M,0})\rho_{c,0}=\\ &=\dfrac{\alpha(\Omega_{M,0})}{2G}(1-\Omega_{M,0}) R_0^3 H_0^2.\nonumber 
\end{eqnarray}
This formula (together with equations~\ref{24b1} and ~\ref{24b2} defining $\alpha(\Omega_{M,0})$) is the solution of the task we consider. We should underline that it disregards the dark energy mass inside $R_0$, which is equal to $4\pi R_0^3\Omega_{\Lambda,0}\rho_{c,0}/3$.

If we accept $\Omega_{M,0}=0.306$ \citep{pdg18}, we obtain for the matter mass inside $R_0$
\begin{equation}
M=2.278\cdot 10^{12} M_\odot\times \left(\dfrac{R_0}{1\text{Mpc}}\right)^3 \left(\dfrac{H_0}{73\frac{\text{km/s}}{\text{Mpc}}}\right)^2
 \label{24a13}
\end{equation}
Thus, the average matter density $\sigma_{M,0}=\alpha \rho_{c,0} \Omega_{\Lambda,0}$ inside $R_0$ does not depend on the group mass: it is universal for all objects with $z\ll 1$. 

The ratio between the average matter density $\sigma_{M,0}$ inside $R_0$ and the dark energy density $\rho_{\Lambda,0}$ is rather large:  $\alpha(\Omega_{M,0}=0.306)\equiv \sigma_{M,0}/\rho_{\Lambda,0} \simeq 5.30$, while $\Omega_{M,0}/\Omega_{\Lambda,0}\simeq 0.441$. Thus, the matter density inside $R_0$ exceeds the average one by more than ten times. Therefore, despite of the fact that the dark energy dominates in the modern Universe, its influence on the Universe dynamics at $r\sim R_0$ is still moderate, though quite noticeable. On the other hand, the presence of the dark energy is quite significant in the following sense: one may see in Figure~\ref{24fig2} that in the hypothetical case of a universe with the same critical density $\rho_{c,0}$ and $\Omega_{M,0}=1$, $\Omega_{\Lambda,0}=0$ (i.e., without dark energy), the mass inside $R_0$ would be $\sim 1.5$ times larger.

Now we may estimate the accuracy of approximate equations for $M(R_0)$ by comparing their results with the exact solution. As an instance, the analytical formula \citep[eqn. 5]{makarov2009} coincides with the exact solution if $\Omega_{M,0}=1$ and overestimates the mass for $\Omega_{M,0}<1$. For equation~(\ref{24a13}) (which corresponds to $\Omega_{M,0}=0.306$), it gives the coefficient $2.348$ instead of the exact value $2.278$. Thus, the difference is almost negligible for the realistic value of $\Omega_{M,0}$, but it rapidly grows as $\Omega_{M,0}\to 0$. In a similar manner, one may estimate the accuracy of other approximate equations and methods.

\textcolor{magenta}{}The preceding consideration assumes that we consider a galaxy group with a negligible redshift $z=0$. However, the result can be used for any $z$ (if $z\ll 1000$, since in the opposite case one may not neglect the radiation). The choice of the 'present moment' is arbitrary in our calculations, and one needs just to find the matter fraction $\Omega_{M,z}$ and the Hubble constant $H_z$ at the moment $z$, and substitute these values instead of $\Omega_{M,0}$ and $H_0$ in (\ref{24b1}), (\ref{24b2}), (\ref{24a12}). Since $a_0/a\equiv (z+1)$, we obtain from (\ref{24a2})
\begin{eqnarray}
\Omega_{M,z}=\dfrac{\Omega_{M,0}(z+1)^3}{\Omega_{M,0}(z+1)^3+\Omega_{\Lambda,0}}\nonumber\\
H^2_z=H^2_0 \left(\Omega_{M,0}(z+1)^3 +\Omega_{\Lambda,0}\right).\label{24c2}
\end{eqnarray}
As a toy example, if $z\gg 1$, we may neglect the dark energy ($\Omega_{\Lambda,z}\simeq 0$, $\Omega_{M,z}\simeq 1$), and simplify (\ref{24a12}) with the help of (\ref{24c1}) and (\ref{24c2}) to
\begin{eqnarray}
M=\dfrac{9 \pi^2}{32 G} R_0^3 H_0^2 \left(\Omega_{M,0}(z+1)^3 +\Omega_{\Lambda,0}\right),\quad \text{if}\quad z\gg 1. \label{24c3}
\end{eqnarray}


\label{lastpage}
\end{document}